\documentclass[a4paper,11pt]{article}
\usepackage{jcappub} 
\usepackage{lineno}

\usepackage{amsmath}
\usepackage{amsfonts}
\usepackage[english]{babel}
\usepackage{amssymb}
\usepackage{subcaption}
\usepackage{multicol}
\usepackage{float}
\usepackage[dvipsnames]{xcolor}
\usepackage{wrapfig}
\usepackage{graphicx}
\usepackage{color}
\usepackage{listings}
\usepackage{xcolor}
\usepackage{hyperref}
\usepackage{caption}
\usepackage{multirow}
\usepackage[mathscr]{euscript}

\hypersetup{
    colorlinks=true,
    linkcolor=red, 
    citecolor=blue, 
    urlcolor=purple 
}

\title{Is there a dynamical tendency in $H_0$ with late time measurements?}
\author[a]{Mauricio Lopez-Hernandez}

\author[a,b]{Josue De-Santiago}
    \affiliation[a]{ Departamento de F\'isica, Centro de Investigaci\'on y de Estudios Avanzados del I.P.N. 
                    Apartado Postal 14-740, 07000, Ciudad de M\'exico, M\'exico}
    \affiliation[b]{Secretar\'ia de Ciencia, Humanidades, Tecnolog\'ia e Innovaci\'on,
    			Av. Insurgentes Sur 1582, 03940, Ciudad de M\'exico, M\'exico}
\date{\today}

\emailAdd{mauricio.lopez@cinvestav.mx}
\emailAdd{Josue.desantiago@cinvestav.mx}

\abstract{The discrepancy between the Hubble constant $H_0$ values derived from early-time and late-time measurements, reaching up to $4\sigma$, represents the most serious challenge in modern cosmology and astrophysics. In this work, we investigate if a similar tension exists between only late time measurements at different redshifts. We use the latest public datasets including Cosmic Chronometers, Megamasers, SNe Ia and DESI-BAO, that span from redshift $z \sim 0$ up to $z\sim 2.3$. By dividing the data into redshift bins, we derive $H_0$ values from each bin separately.
Our analysis reveals a phenomenological dynamic evolution in $H_0$ across different redshift ranges, with a significance from $1.5\sigma$ and $2.3\sigma$, depending on the parameterization.
Consistency of the model demands observational constancy of $H_0$ since it is an integration constant within the Friedmann-Lema\^itre-Robertson-Walker (FLRW) metric. Thus, these findings suggest that the observed Hubble tension might not only exist between early and late-time measurements but also among late-time data themselves, providing new insights into the nature of the Hubble tension.}

\begin{document}
\maketitle
\flushbottom

\section{\label{sec:intro}Introduction}

Our current understanding of the Universe relies on the $\Lambda$CDM model, which incorporates the cosmological constant $\Lambda$ and a cold dark matter component. This model is the most accepted and successful in explaining most cosmological observations \cite{refId0,eBOSSS}. One of the most important challenges within the astrophysics and cosmology community is the discrepancy of more than $4\sigma$ for the Hubble constant $H_0$, between local measurements using the cosmic distance ladder and early-time measurements of the cosmic microwave background (CMB).  The former was based on direct observations of Supernovae Type Ia (SNe Ia) calibrated with Cepheids by the SH0ES team \cite{Riess_2022} and the latter was obtained by the Planck satellite, which relies on the $\Lambda$CDM model \cite{refId0}. This so-called \textbf{Hubble tension} has led to an exhaustive search for potential explanations. A wide range of models have been used to address the $H_0$ tension, proposing either a dark energy (DE) component during the early evolution of the Universe \cite{PhysRevLett.124.161301,Gogoi_2021,PhysRevD.103.043518,PhysRevD.101.063523,PhysRevD.103.063502,Chudaykin_2020}, a DE component with a time-varying equation of state \cite{Wang_2018,PhysRevD.99.043543,Guo_2019}, extra interactions between the components of the Universe \cite{PhysRevD.62.043511,PhysRevD.98.043521,PhysRevD.97.103530,PhysRevD.101.123505}, or modified gravity \cite{PhysRevD.101.103505,refId0Late-time,PhysRevD.102.124029}. However, in most solutions, reducing the tension leads to a mismatch with other well-measured cosmological quantities or fails to fully address the Hubble tension, so no proposal so far presents solid arguments to claim it is better than all the others \cite{Jedamzik2021-sg,PhysRevD.104.063524,Mortsell_2018,PhysRevD.99.043514,Arendse_Cosmic_dissonance,PhysRevD.101.043533,PhysRevD.102.123515,10.1093/mnras,PhysRevD.97.103511,Lin_2021}. For a more comprehensive study about the $H_0$ tension and alternative proposals, see the reviews 
\cite{Di_Valentino_2021,Universe9120501,PERIVOLAROPOULOS2022101659}.

In the $\Lambda$CDM model, $H_0$ is a constant by definition: it represents the Hubble parameter $H(z)$ at the present epoch $H_0=H(z=0)$. However, recent results have shown that $H_0$ depends on the redshift of the probes used to determine it. The first of these results, from six gravitationally lensed quasars with measured time delays by the H0LiCOW program, indicated that $H_0$ decreases with lens redshift \cite{H0LiCOW}, with no evidence suggesting that this trend is due to unaccounted systematics \cite{H0LiCOW_sistematicos}. Subsequent studies have found a similar feature in different datasets \cite{PhysRevD.102.103525,OCOLGAIN2024101464,Malekjani2024,Dainotti_2021,10.1093/mnras/stac2728,evidence_decreasing,Jia_2025,galaxies10010024,PhysRevD.106.L041301,PhysRevD.103.103509}, offering a new perspective on the Hubble tension due to an apparent redshift dependence of a quantity that should, tautologically, be a constant.

Recently, the Dark Energy Spectroscopic Instrument (DESI) collaboration published its cosmological constraints from the first year of observations. The results favor models with a DE component whose equation of state varies with time by more than $2.6\sigma$ \cite{DESI} for the baryon acoustic oscillations (BAO) measurements with similar results from the more recent full shape analysis \cite{adame2024desi} with a significance from $2.5\sigma$ up to $3.8\sigma$ depending on the Supernova data used. This challenges the paradigm of what we previously believed to be correct in the $\Lambda$CDM model. This dynamical DE could be responsible for a $H_0$ that depends on the redshift.

Ref. \cite{PhysRevD.102.103525} found that splitting cosmological data in different redshift bins up to $z=0.7$ gives different estimates to $H_0$ which are in contradiction with a single constant at $2.1\sigma$ level. This leads to the conclusion that the Hubble tension is present even between late-time measurements themselves when taken at different redshifts. Ref.  \cite{PhysRevD.102.103525} assumed a cosmological constant and used BAO-BOSS data \cite{eBOSSS}. The new DESI results lead us to question whether the new data together with a dynamical DE could solve this tension. 

Clearly, there is a need for model-independent techniques to reconstruct the expansion rate of the universe. Cosmography \cite{Visser2005-ko, shafieloo2012gaussian, ruiz2022model} is one of these approaches; since it only uses kinematic parameters together with the assumption of homogeneity and isotropy on large scales, it turns out to be a very useful tool to deal with degeneracies between cosmological models beyond $\Lambda$CDM and to understand the kinematics of our local universe. Letting the data constrain the Hubble rate without dealing with the problems related to the nature of dark energy or dark matter leads to an excellent consistency test to break the assumptions that were once established given the extremely limited data available for cosmology at that time. Then, after reconstructing $H(z)$ with the latest cosmological data we can ask if our cosmological models are consistent with it.

The paper is organized as follows. In Section \ref{sec:theory}, we detail the implemented methodology, along with the theory behind the cosmological probes used in this paper. In Section \ref{sec:data}, we mention the data used. Section \ref{sec:fitting} explains the criteria used to split the data into bins. In Section \ref{sec:results}, we present our results on constraining the value of $H_0$ in each of the bins. Finally, the summary and conclusions can be found in Section \ref{sec:conclu}.

\section{\label{sec:theory} Cosmology}

In light of the new discoveries for the first cosmological results from DESI \cite{DESI}, in this work we use a flat Universe with a dynamical Dark Energy component, with a Chevallier-Polarski-Linder (CPL) equation of state given by $w(a)=w_0 + w_a (1-a)$ \cite{doi:10.1142/S0218271801000822,PhysRevLett.90.091301}. The Friedmann equation is given by the following parameterization 
\begin{equation}
H(z)= H_0 \left[ \Omega_m (1+z)^3 + (1-\Omega_m)f(z)\right]^{1/2},
\label{eq: ecu de Friedmann wowaCDM}
\end{equation}
where
\begin{equation}
f(z)=(1+z)^{3(1+w_0 + w_a)} e^{-3 w_a z/(1+z)},
\label{eq: f(z)}
\end{equation}
and with $H_0$, $\Omega_m$, $w_0$, $w_a$ the free parameters. 

Following \cite{PhysRevD.102.103525}, we will split different cosmological data according to their redshift in order to estimate the Hubble constant using only data in certain redshift ranges. This method involves binning a modern dataset that includes the new DESI data, cosmic chronometers (CC), megamasers, and SNe Ia samples. The SNe Ia samples comprise the Pantheon+ compilation, the new Union3 compilation, and Dark Energy Survey (DES) Year 5 data release. With these new data, we were able to expand the redshift range in the analysis to observe in more depth the evolution of $H_0$ with the redshift of the data.

Cosmic chronometers provide a direct estimate of the expansion rate of the Universe. This method takes advantage of the fact that the Hubble parameter, $H(z)$, can be directly expressed as a function of the time differential of the Universe, $dt$, over a given redshift interval, $dz$ \cite{Jimenez_2002}
\begin{equation}
H(z) = - \frac{1}{1+z} \frac{dz}{dt}. 
\label{eq: H(z) para CC}
\end{equation}
CCs evolve homogeneously as a function of cosmic time and on a timescale much longer than their age difference. Given $H(z)$, to obtain $H_0$ from eq. \eqref{eq: ecu de Friedmann wowaCDM} it is necessary to determine $\Omega_m$, $w_0$, $w_a$ from other observations.

The water megamasers found in the accretion disks of supermassive black holes in active galactic nuclei (AGN) act like a laser beam in the microwave band, providing a method to calculate extragalactic distances without relying on the cosmic distance ladder \cite{Pesce_2020}. The disk model returns an estimate of the angular diameter distance $\hat{D}_\text{A}$ to the galaxy and its recession velocity $\hat{v}_i$. The decomposition of this velocity between its peculiar and cosmological components $\hat{v}_i = v_\textrm{pec,i}+cz_i$ is unknown for objects this close. Each $\hat{D}_\text{A}$ is related to its expected redshift $z_i$ and the cosmological parameters by the equation
\begin{equation}
D_{\textrm{A}} (z) = \frac{c}{1+z} \int_{0}^{z} \frac{dz^{\prime}}{H(z^{\prime})}. 
\label{eq: Distancia angular}
\end{equation}
Following \cite{Megamaserproject}, we take peculiar velocities of the galaxies into account incorporating $\sigma_{\textrm{pec}}=250$ km $\textrm{s}^{-1}$ into the velocity uncertainties. The chi-squared used to constrain the cosmological parameters is then given as 
\begin{equation}
\chi^{2} = \sum^{N}_{i=1} \left[ \frac{(v_i-\hat{v}_i)^{2}}{\sigma_{v,i}^{2}+\sigma_{\textrm{pec}}^{2}} + \frac{(D_\textrm{A} (v_i/c)-\hat{D}_i)^{2}}{\sigma_{D,i}^{2}} \right],
\label{eq: chi2 de masers}
\end{equation}
where the expected cosmological recession velocities are given by $v_i = c z_i$, $N$ the number of data, $\sigma_{v,i}$ is the statistical uncertainty in the velocity measurement $\hat{v}_i$, and $\sigma_{D,i}$ is the standard deviation of the distance measurement $\hat{D}_i$. We take the expected velocities $v_i$ for megamaser host galaxies as nuisance parameters.

The SNe Ia data give the apparent magnitude $m_\textrm{b}$ of each SNe, which can be compared with the predicted model given by
\begin{equation}
m_\textrm{b} = \textrm{M} + \mu(z) = \textrm{M} + 25 + 5\textrm{log}_{10} \left( \frac{D_\textrm{L} (z)}{\textrm{Mpc}} \right),
\label{eq: magnitud aparente}
\end{equation}
where M is the absolute magnitude for SNe Ia, and $D_\textrm{L} (z)$ is the luminosity distance, which in the case of SNe Ia can be written, according to \cite{Kenworthy_2019}, by
\begin{equation}
D_\textrm{L} (z)= (1+z_{\textrm{hel}})(1+z_{\textrm{cmb}}) D_{\textrm{A}} (z_{\textrm{cmb}})
\label{eq: distancia luminica},
\end{equation}
$z_{\textrm{hel}}$ is the heliocentric redshift of the SNe (or its host galaxy), and $z_{\textrm{cmb}}$ is this redshift corrected for the peculiar motion of the solar system with respect to the CMB. Note that the parameters M and $H_0$ are degenerate when analyzing SNe alone. Therefore, SNe measurements are used to constrain other cosmological parameters, particularly the matter density $\Omega_m$.

The BAO measurements provide distances relative to the sound horizon, $r_d$, as a function of redshift \cite{eBOSSS}. Along the line-of-sight direction, a measurement of the redshift interval, $\Delta z$, gives a means to measure the Hubble distance at redshift $z$,
\begin{equation}
D_{\textrm{H}} (z) = \frac{c}{H(z)}.
\label{eq: DH}    
\end{equation}
Along the transverse direction, measuring the angle $\Delta \theta$ subtended by the BAO feature at $z$ estimates the comoving angular diameter distance, $D_\text{M} (z)$
\begin{equation}
D_{\textrm{M}} (z) = (1+z) D_{\textrm{A}}(z),
\label{eq: Distancia DM}
\end{equation}
These measurements can also be summarized by a single estimate representing the spherically averaged distance
\begin{equation}
D_{\textrm{V}} (z) = \left[ z D^{2}_{\textrm{M}} (z) D_{\textrm{H}} (z) \right]^{1/3},
\label{eq: DV}
\end{equation}
BAO measurements constrain these distances all divided by the sound horizon $r_{\textrm{d}}$. Therefore, $H_0$ and $r_{\textrm{d}}$ are degenerate with each other. BAO measurements constrain the product $H_0 r_{\textrm{d}}$, and therefore are key in the treatment of Hubble tension. This constraint implies that a higher $H_0$, as that obtained by SH0ES \cite{Riess_2022}, requires a lower $r_{\text{d}}$, which is incompatible with the value inferred by the Planck Collaboration \cite{refId0}. In this work, we will not use the CMB prior on $r_{\textrm{d}}$ as this will imply the use of data outside our redshift bins.

\section{\label{sec:data} Data}
To constrain $H_0$, we use the following public data up to the redshift $z < 2.33$, including their systematic errors and covariance matrices. These data correspond to the latest public measurements of the observables from the previous section:

\begin{enumerate}
\renewcommand{\theenumi}{\Roman{enumi}}
    \item{Distances and recession velocities from 6 megamasers in the range $0.002 \leq z \leq 0.034$, from the Megamaser Cosmology Project \cite{Pesce_2020,Megamaserproject,Reid_2019}.}
    
    \item{32 cosmic chronometers data in the range $0.07 \leq z \leq 2$, from refs. \cite{Zhang_2014,Jimenez_2003,PhysRevD.71.123001,M.Moresco_2012,Moresco_2016,10.1093/mnras/stx301,Daniel_Stern_2010,Borghi_2022,10.1093/mnrasl/slv037}. We estimate the covariance matrix\footnote{The detailed recipe on how to correctly estimate the covariance matrix for cosmic chronometers is given in \url{https://gitlab.com/mmoresco/CCcovariance}} between these data points following ref. \cite{Moresco_2020}.}
    
    \item{BAO measurements in galaxy, quasar and Lyman-$\alpha$ forest tracers from the Year 1 data release of DESI. These are measurements of the transverse comoving distance $D_{\textrm{M}}$ and Hubble rate $D_\textrm{H}$, or their combination $D_{\textrm{V}}$, relative to the sound horizon $r_{\textrm{d}}$ with the following effective redshifts $z_{\textrm{eff}}=0.30, 0.51, 0.71, 0.93, 1.32, 1.49, 2.33$ \cite{DESI}.}
    
    \item{For the superenovae  type Ia measurements, we use any of the three latest compilations. 1,590 SNe Ia from the Pantheon+ compilation in the redshift range $0.01 \leq z \leq 2.26$ \cite{Brout_2022}. Recently published results from the Year 5 data release of the Dark Energy Survey (DES Y5) that include 1,635 new photometrically classified SNe Ia and 194 low-redshift SNe Ia (which are also common to Pantheon+), spanning $0.025 \leq z \leq 1.3$ \cite{Abbott_2024}. Finally, 22 binned data points from 2087 SNe Ia from the Union3 sample in the range $0.05 \leq z \leq 2.26$ \cite{rubin2023union}, which has 1360 SNe in common with Pantheon+. We include both the statistical and systematic uncertainties in the three sets. We emphasize that these SNe Ia data are not independent of each other and therefore we cannot combine them. However, they differ in the way they analyze systematic errors and astrophysical parameters, making it interesting to compare the results of each of them.}
\end{enumerate}

Compared with the data used in ref. \cite{PhysRevD.102.103525} which includes the BOSS BAO data \cite{eBOSSS} and the Pantheon sample \cite{Pantheon_old}, we have the BAO data from DESI Y1, the recent supernovae compilations Pantheon+, Union3, and DES Y5, as well as three extra measurements of CC.

Since the Pantheon+, DES, and Union3 data cannot be combined, we present each of their results separately. The analyzes always include the megamasers+CC+DESI data, which we will call ``Base'', but differ in the SNe sample. We denote each case as Base+PantheonPlus, Base+Union3 and Base+DES.

\section{\label{sec:fitting} Fitting}

\begin{table}
\centering
\resizebox{9.5 cm}{!} {
\renewcommand{\arraystretch}{1.5}
\begin{tabular}{c|c|c|c}
\hline
\multicolumn{4}{c}{Base+PantheonPlus} \\ \hline
Bin & Data & Range & $\bar{z}$ \\ \hline
1 & Megamasers, SNe & $0.01 < z \leq 0.069$ &  $0.032$ \\ 
2 & SNe, CC & $0.069 < z \leq 0.199$ & $0.15$ \\
3 & SNe, CC, DESI-BGS & $0.199 < z \leq 0.425$ & $0.30$ \\
4 & SNe, CC, DESI-LRG1 ($z=0.51$) & $0.425 < z \leq 0.625$ & $0.51$ \\
5 & SNe, CC, DESI-LRG2 ($z=0.71$) & $0.625 < z \leq 0.7891$ & $0.71$ \\
6 & SNe, CC, DESI-LRG3+ELG1 & $0.7891 < z \leq 1.13$ & $0.93$ \\
7 & SNe, CC, DESI-ELG2, DESI-QSO & $1.13 < z \leq 1.65$ & $1.38$ \\
8 & SNe, CC, DESI-Lya & $1.65 <z \leq 2.3$ & $1.94$ \\ \hline
\multicolumn{4}{c}{Base+Union3} \\ \hline
Bin & Data & Range & $\bar{z}$ \\ \hline
1 & Megamasers, SNe & $0 < z \leq 0.069$ &  $0.05$ \\ 
2 & SNe, CC & $0.069 < z \leq 0.19$ & $0.12$ \\
3 & SNe, CC, DESI-BGS & $0.19 < z \leq 0.44$ & $0.30$ \\
4 & SNe, CC, DESI-LRG1 ($z=0.51$) & $0.44 < z \leq 0.649$ & $0.51$ \\
5 & SNe, CC, DESI-LRG2 ($z=0.71$) & $0.649 < z \leq 0.79$ & $0.71$ \\
6 & SNe, CC, DESI-LRG3+ELG1 & $0.79 < z \leq 1.23$ & $0.93$ \\
7 & SNe, CC, DESI-ELG2, DESI-QSO & $1.23 < z \leq 1.7$ & $1.36$ \\
8 & SNe, CC, DESI-Lya & $1.7 < z \leq 2.3$ & $2.26$ \\ \hline
\multicolumn{4}{c}{Base+DES} \\ \hline
Bin & Data & Range & $\bar{z}$ \\ \hline
1 & Megamasers, SNe & $0 < z \leq 0.069$ &  $0.041$ \\ 
2 & SNe, CC & $0.069 < z \leq 0.19$ & $0.12$ \\
3 & SNe, CC, DESI-BGS & $0.19 < z \leq 0.401$ & $0.30$ \\
4 & SNe, CC, DESI-LRG1 ($z=0.51$) & $0.401 < z \leq 0.63$ & $0.51$ \\
5 & SNe, CC, DESI-LRG2 ($z=0.71$) & $0.63 < z \leq 0.826$ & $0.71$ \\
6 & SNe, CC, DESI-LRG3+ELG1 & $0.826 < z \leq 1.13$ & $0.93$ \\ \hline
\end{tabular}
}
\caption{\footnotesize{Summary of the data, range in redshift and $\bar{z}$ in each bin for each different dataset.}}
\label{tab: bines}
\end{table}

In this section, we describe how we fit the cosmological parameters using the data specified in the last section. The code used in this section is public at \footnote{See \url{https://github.com/MauLoHdz/LUDB}}. Our objective is to see if the estimation of the Hubble constant $H_0$ is consistent when using data at different redshifts. We will split the total redshift range from our data into separate redshift bins with $z \in (z^{i}_{min},z^{i}_{max})$, and we use only data coming from that bin to estimate $H_0$.

The redshift widths of the bins vary between each other, as we follow the strategy used in \cite{PhysRevD.102.103525} requiring that the average redshift of each bin coincides between megamasers, SNe and BAO. We use the weighted average $\bar{z}$ given by 
\begin{equation}
\bar{z_i} = \frac{\sum_{k}^{N_i} z_{k} (\sigma_k)^{-2}}{\sum_{k}^{N_i} (\sigma_k)^{-2}} \,,
\label{eq: z pesado}
\end{equation}
where $N_i$ denotes the number of data used in the bin $i$ and $\sigma_k$ denotes the error in the observable at redshift $z_k$. For megamasers, we use $\sigma_{k}^{2}=\sigma_{\hat{v},k}^{2}+\sigma_{\textrm{pec}}^{2}+\sigma_{\hat{D},k}^{2}$. The bins constructed in each analysis are summarized in table \ref{tab: bines}.

Not all bins contain all the data. The megamasers appear only in the first redshift bin, while some bins lack DESI data. We only require that all bins contain either CC or megamasers data to break the $H_0$-$r_d$ and $H_0$-M degeneracies, mentioned in section \ref{sec:theory}. We emphasize that the condition \eqref{eq: z pesado} leads to different binnings and different $\bar{z}$'s depending on the SNe's used (PantheonPlus, Union3 and DES).

To fit the parameters, we used the Python module \textit{emcee} \cite{emcee}. We set flat priors for all the parameters,  in particular for $H_0$ the prior is large enough that its fitting is mainly influenced by the measurements
\begin{equation}
H_0 \in (0,150).
\end{equation}

On the other hand, the data contained in a single bin lack the power to constrain the equation of state parameters $w_0$ and $w_a$. Therefore, the posteriors of these parameters are mainly influenced by their priors. We considered large prior ranges
\begin{equation}
w_a \in (-2.5,2.5), \; \; w_0 \in (-1.5,-1/3),
\end{equation}
to allow for enough freedom in the DE dynamics. These ranges contain both the DESI estimation\footnote{We recall that $w_0=-0.64 \pm 0.11$, and $w_a=-1.27^{+0.40}_{-0.34}$ for DESI+CMB+Union3 \cite{DESI}.} and the cosmological constant $w_0=-1$, $w_a=0$.
After we fit the parameters, $w_0$ and $w_a$ take values in their entire allowed region without preference for a particular value. This comes from the narrow ranges in $z$ for each bin which limit the evolution of $f(z)$ in equation \eqref{eq: f(z)} and therefore the possible observable effects from a dynamical Dark Energy. In subsection \ref{sec:unbinned} we do find limits for $w_0$ and $w_a$ using the full redshift range of the data.

Not all bins fit all parameters, since each bin may or may not contain a certain type of data. The only parameters fitted in all bins are $\Omega_m$, $H_0$, $w_0$, $w_a$, and the absolute magnitude of SNe Ia M.

\section{\label{sec:results} Results}

\begin{figure}
\centering
\includegraphics[scale=0.5]{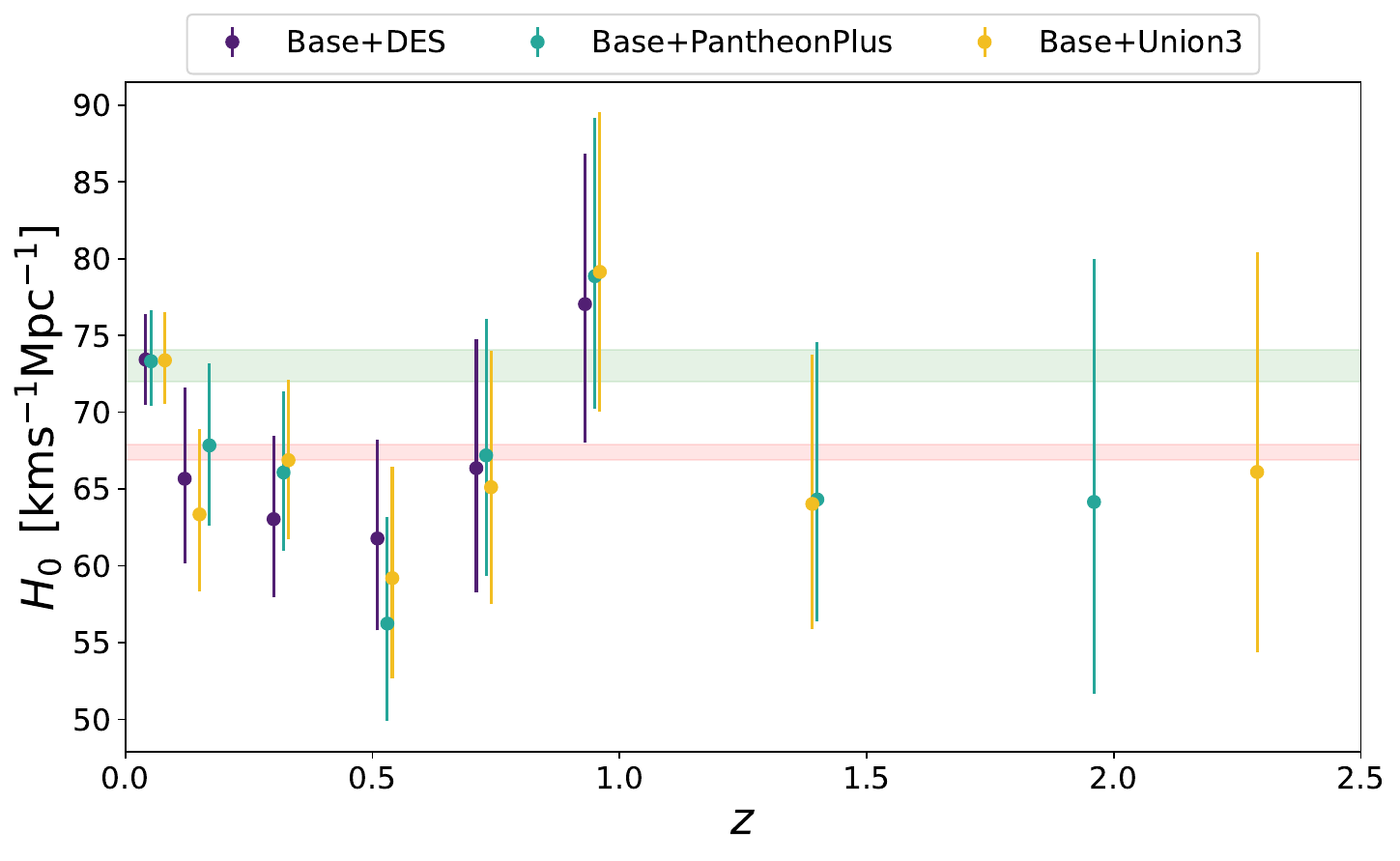}
\caption{\footnotesize{Best-fit values for the Hubble constant in each bin and for each different dataset. The Base+DES data cover the range $z \leq 1.3$, so it includes only 6 redshift bins, compared to Base+PantheonPlus and Base+Union3, which cover a wider range. For a better visibility, we shifted the redshift centers $\bar{z}_i+0.02$ for Base+PantheonPlus and $\bar{z}_i+0.03$ for Base+Union3. The centers of the bins $\bar{z}_i$ differ for each dataset due to eq. \eqref{eq: z pesado}. The
green horizontal band corresponds to the value from SH0ES \cite{Riess_2022} ($H_0=73.04 \pm 1.04$ km $\text{s}^{-1}$$ \text{Mpc}^{-1}$) and the light red horizontal band corresponds to
the value from Planck 2018 \cite{refId0} within a $\Lambda$CDM scenario ($H_0=67.4 \pm 0.5$ km $\text{s}^{-1}$$ \text{Mpc}^{-1}$). }}
\label{fig:all_results_H0}
\end{figure}

We show the fitted values of $H_0$
for each bin in Table \ref{tab:resultados H0 de cada bin} and illustrate them in Figure \ref{fig:all_results_H0}. In all three analyses, $H_0$ decreases as a function of the redshift for $z < 0.6$, which is consistent with \cite{PhysRevD.102.103525}. In the next two bins, the value of $H_0$ increases to a maximum of $\sim 78$ km $\text{s}^{-1}$ $\text{Mpc}^{-1}$ and then decreases again. Note that the scarcity and low quality of the data inflate the errors at high-redshifts.

\begin{table*}[t!] 
\centering
\resizebox{\columnwidth}{!} {
    \renewcommand{\arraystretch}{2}
    \begin{tabular}{c c c c c c c c c}
    \hline
    \multirow{2}{*}{\shortstack{Data}} & \multicolumn{8}{c}{$H_0 \left[ \frac{\text{km}}{\text{s Mpc}} \right]$} \\ 
       & Bin 1 & Bin 2 & Bin 3 & Bin 4 & Bin 5 & Bin 6 & Bin 7 & Bin 8 \\ \hline
    Base+PantheonPlus & $73.32^{+3.34}_{-2.92}$ & $67.84^{+5.37}_{-5.20}$ & $66.08^{+5.26}_{-5.11}$ & $56.24^{+6.91}_{-6.36}$ & $67.19^{+8.92}_{-7.82}$ & $78.86^{+10.30}_{-8.62}$ & $64.32^{+10.24}_{-7.91}$ & $64.16^{+15.82}_{-12.51}$  \\ 
    Base+Union3 & $73.38^{+3.12}_{-2.82}$ & $63.35^{+5.58}_{-5.03}$ & $66.88^{+5.26}_{-5.17}$ & $59.20^{+7.26}_{-6.53}$ & $65.12^{+8.86}_{-7.63}$ & $79.14^{+10.38}_{-9.13}$ & $64.03^{+9.69}_{-8.14}$ & $66.11^{+14.32}_{-11.77}$ \\ 
    Base+DES & $73.43 \pm 2.97$ & $65.67^{+5.97}_{-5.52}$ & $63.04^{+5.41}_{-5.11}$ & $61.78^{+6.45}_{-5.99}$ & $66.36^{+8.37}_{-8.09}$ & $77.04^{+9.79}_{-9.02}$ & -- & -- \\ \hline
    \end{tabular}
    }
\caption{\footnotesize{Numerical values for the best-fit $H_0$ in each bin and for each dataset. To see the redshift range of each of the bins refer to table \ref{tab: bines}. This data is plotted in fig. \ref{fig:all_results_H0}.}}
\label{tab:resultados H0 de cada bin}
\end{table*}

In Figure \ref{fig:all_results_H0} we see that the estimations in $H_0$ seem to depend on the redshift. In order to quantify how much this measurements deviate from a constant, we introduce different parameterizations
\begin{equation}
\widetilde H_0 (\bar z,\theta_k) \,,
\label{eq: general_parametrization}
\end{equation}
that are functions of the redshift of the measurements $\bar z$ and some parameters $\theta_k$ for $k=1,2,\cdots$ which allow $\widetilde H_0 $ to deviate from a constant.
Assuming statistical independence we compute the total posterior as
\begin{equation}
\text{P}_{\text{tot}} (\theta_k) = \prod_{i=1}^{i_{max}} \text{P}_i \left(\widetilde H_0(\bar{z}_i,\theta_k)\right) \,,
\label{eq: log de la P_total}
\end{equation}
where the posteriors $P_i$ result from the fits of each of the bins. With this, we fit the parameters $\theta_k$ and estimate how much $\widetilde H_0$ deviates from a constant.

We evaluated how statistically significant the deviation from a constant $H_0$ is using three different methods. First, we use the Bayesian evidence given by \cite{Trotta:2008qt}
\begin{equation}
E = \int \Pi(\theta_k) \mathscr{L}(\theta_k) d\theta_k \,,
\end{equation}
where $\Pi$ is the prior and $\mathscr{L}$ the Likelihood of the parameters $\theta_k$. We compute the integral using the nested integration algorithm, Polychord \cite{skilling2006nested, Handley:2015fda}. Here, a larger value means a model with more probability. We report the Bayes factor given by
\begin{equation}
B = \frac{E_{\text{var}}}{E_{\text{cons}}}
\end{equation}
and use it to determine if either the variable model or the constant is favored by the data.

Secondly, the Akaike Information Criterion (AIC) defined as \cite{AIC_Akaike,AIC2}
\begin{equation}
\text{AIC} = -2 \log \mathscr{L}_{\text{max}} + 2k,
\label{eq: AIC}
\end{equation}
where $\mathscr{L}_{\text{max}}$ is the maximum likelihood that can be obtained within the model and $k$ the number of free parameters. Between the two models, variable and constant, the one with the lowest AIC value is preferred by the data. This criterion favors the larger $\mathscr{L}_{\text{max}}$, but penalizes too many extra parameters. In the next subsections, we report $\Delta\text{AIC} =  \text{AIC}_{\text{var}} - \text{AIC}_{\text{cons}}$.

Thirdly, we determine the number of standard deviations ($\sigma$) from a constant. For parameterizations with a single extra parameter, we only divide the distance to a constant by the standard deviation of the parameter. For models with more parameters, we compute the $p$-value of the variable model and, assuming a Gaussian distribution, we determine the number of standard deviations associated with that $p$-value. This is the least rigorous of our tests, but serves to communicate the relevance of our results as the cosmological community is familiar with the concept that a few $\sigma$'s can be the result of statistical fluctuations and not a physical effect.

\subsection{\label{subsec:quadratic paramete}Quadratic parameterization}

The first 6 bins in Figure \ref{fig:all_results_H0} seem to indicate a quadratic behaviour, therefore we propose a quadratic parameterization of the form
\begin{equation}
\widetilde H_0 (z) = \hat{H}_0 +mz + bz^{2},
\label{eq: Función cuadrática}
\end{equation}
with parameters $\hat{H}_0$, $m$ and $b$. For the fit, we exclude the results for bins 7 and 8 in order to avoid biases caused by the abrupt drop in the value of $H_0$ in bin 7, and the large error bars in bin 8.

The results are shown in Table \ref{tab:resultados cuadrática} and Figure \ref{fig:ajuste_cuadratico}, which shows the best fit to this quadratic function. We also drew the 2.3, 16, 84 and 97.7 percentiles of the fitted samples as a function of redshift to represent the 1$\sigma$ and 2$\sigma$ errors.

From Table \ref{tab:resultados cuadrática} we can see that the parameters $m$ and $b$ deviate from zero, favoring a non-constant $H_0$ model. However, the values of the Bayes factor and the $\Delta$AIC show that this parameterization very modestly improves the fit to the data given by the constant $H_0$. Only accounting for a barely worth mentioning result according to the Jeffrey's scale \cite{jeffreys1998theory} of the Bayes factor. And according to the $\Delta$AIC scale in \cite{AIC2}, which corresponds to no preference between the models.

It is evident that the low values of $H_0$ in the last two bins can lead to a fitting that is compatible with a constant $H_0$ ($m=0$ and $b=0$). This can be considered a limitation of the quadratic parameterization, which asks for $H_0$ to grow up to infinity at large redshifts. As this parameterization is purely phenomenological, we replace it with one that will not exhibit this behavior in the next subsection.

\begin{table*}[t]
    \renewcommand{\arraystretch}{1.7}
    \centering
    \begin{tabular}{c c c c c c c}
    \hline
    Data & $\hat{H}_{0} \left[ \frac{\text{km}}{\text{s Mpc}} \right]$ & $m \left[ \frac{\text{km}}{\text{s Mpc}} \right]$ & $b \left[ \frac{\text{km}}{\text{s Mpc}} \right]$ & $\sigma$ & $\Delta$AIC & $B$\\ \hline
    Base+PantheonPlus & $75.18^{+3.76}_{-3.42}$ & $-65.96^{+26.31}_{-26.69}$ & $72.27^{+31.63}_{-30.81}$ & 1.5 & -1.67 & 1.59 \\ 
    Base+Union3 & $75.06^{+3.79}_{-3.47}$ & $-63.98^{+29.04}_{-28.90}$ & $70.32^{+34.48}_{-33.89}$ & 1.3 & -0.70 & 1.30 \\ 
    Base+DES & $74.64^{+3.62}_{-3.36}$ & $-64.89^{+26.68}_{-26.49}$ & $71.83^{+32.00}_{-32.08}$ & 1.5 & -1.52 & 1.72 \\ \hline
    \end{tabular}
    \caption{\footnotesize{Constraints on the parameters $\hat{H}_0$, $m$ and $b$ in the quadratic parameterization \eqref{eq: Función cuadrática} for the first 6 bins. We compared the model to a constant $H_0$ and determine the Bayes factor $B$, $\Delta$AIC and $\sigma$ deviation. While the $\sigma$ values favors a dynamic $H_0$, the Bayes factor indicates only a barely worth mentioning preference for the quadratic model, and the $\Delta$AIC shows no preference between them, as determined using the scale provided by \cite{AIC2}.}}
    \label{tab:resultados cuadrática}
\end{table*}

\subsection{Fourier parameterization}

If we consider all the bins in Figure \ref{fig:all_results_H0}, we see a possible oscillatory behavior. Therefore, we follow ref. \cite{Tamayo} by proposing a parameterization in terms of the Fourier series
\begin{equation}
\widetilde H_0 (z,b_k,c_k) = \hat H_0 + 
\sum_k \left[ b_k \sin (2k\pi a) + 
c_k \cos (2k\pi a) \right],
\end{equation}
where $a=1/(1+z)$ is the scale factor. After fitting different cuts of the series, we have obtained that the second term in the cosine series $c_2$ follows the oscillatory trend particularly well. In order to simplify the parameterization we stay with
\begin{equation}
\widetilde H_0 (z) = \hat{H}_0 + A\cos(4\pi a).
\label{eq: modelo cos(2x)}
\end{equation}
with only the two parameters $\Hat{H}_0$ and $A$.

\begin{figure}
\begin{minipage}[c]{0.485\linewidth}
\includegraphics[width=1\linewidth]{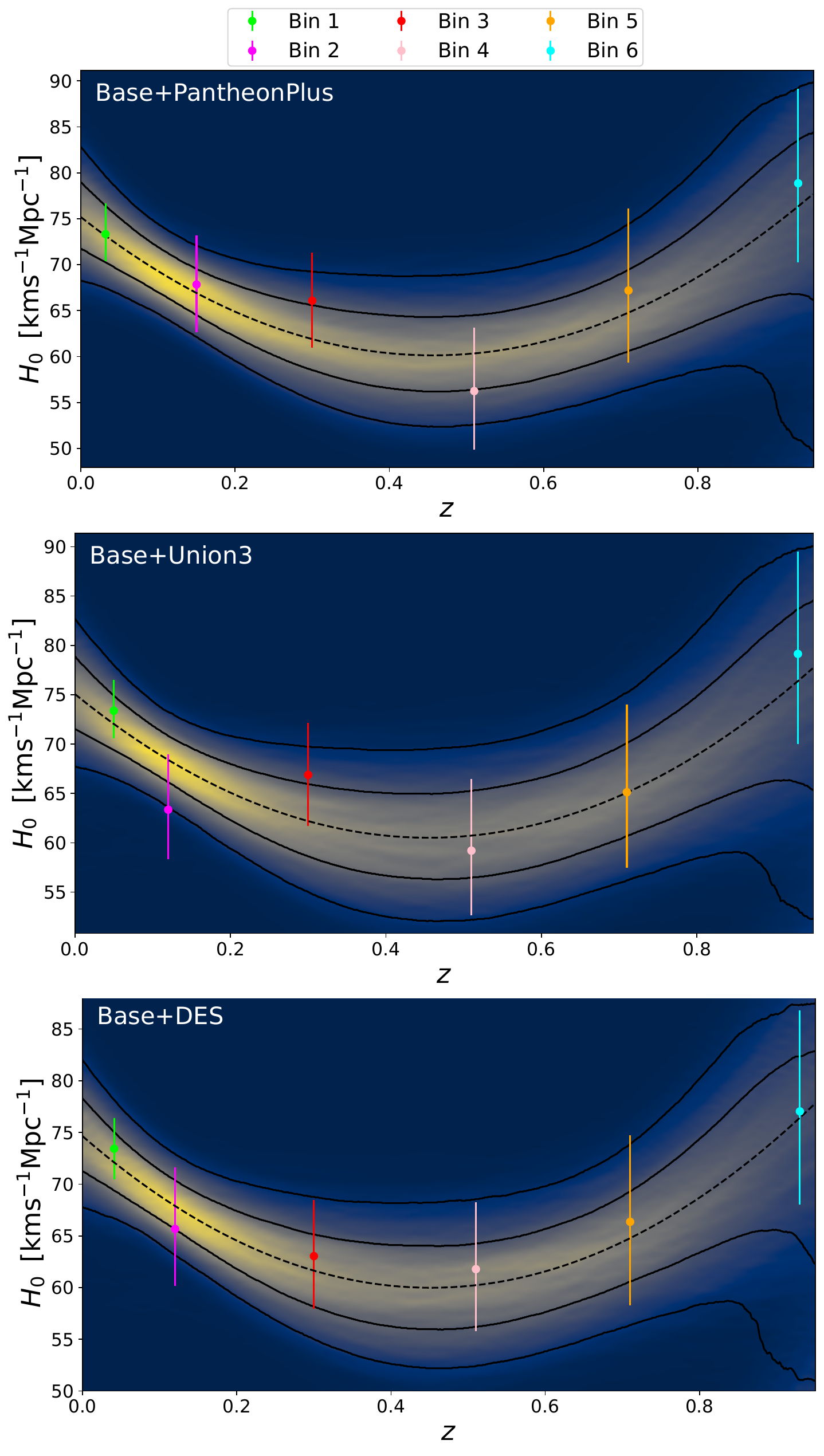}
\caption{\footnotesize{The fit of the quadratic function \eqref{eq: Función cuadrática} to the binned results for $H_0$ in the Base+PantheonPlus sample, Base+Union3 sample and Base+DES sample. The dashed line represents the best fit, while the external solid lines represent the 1$\sigma$ and 2$\sigma$ errors. For this parametrization we only use the first 6 bins of each dataset.}}
\label{fig:ajuste_cuadratico}
\end{minipage}
\hfill
\begin{minipage}[c]{0.485\linewidth}
\includegraphics[width=\linewidth]{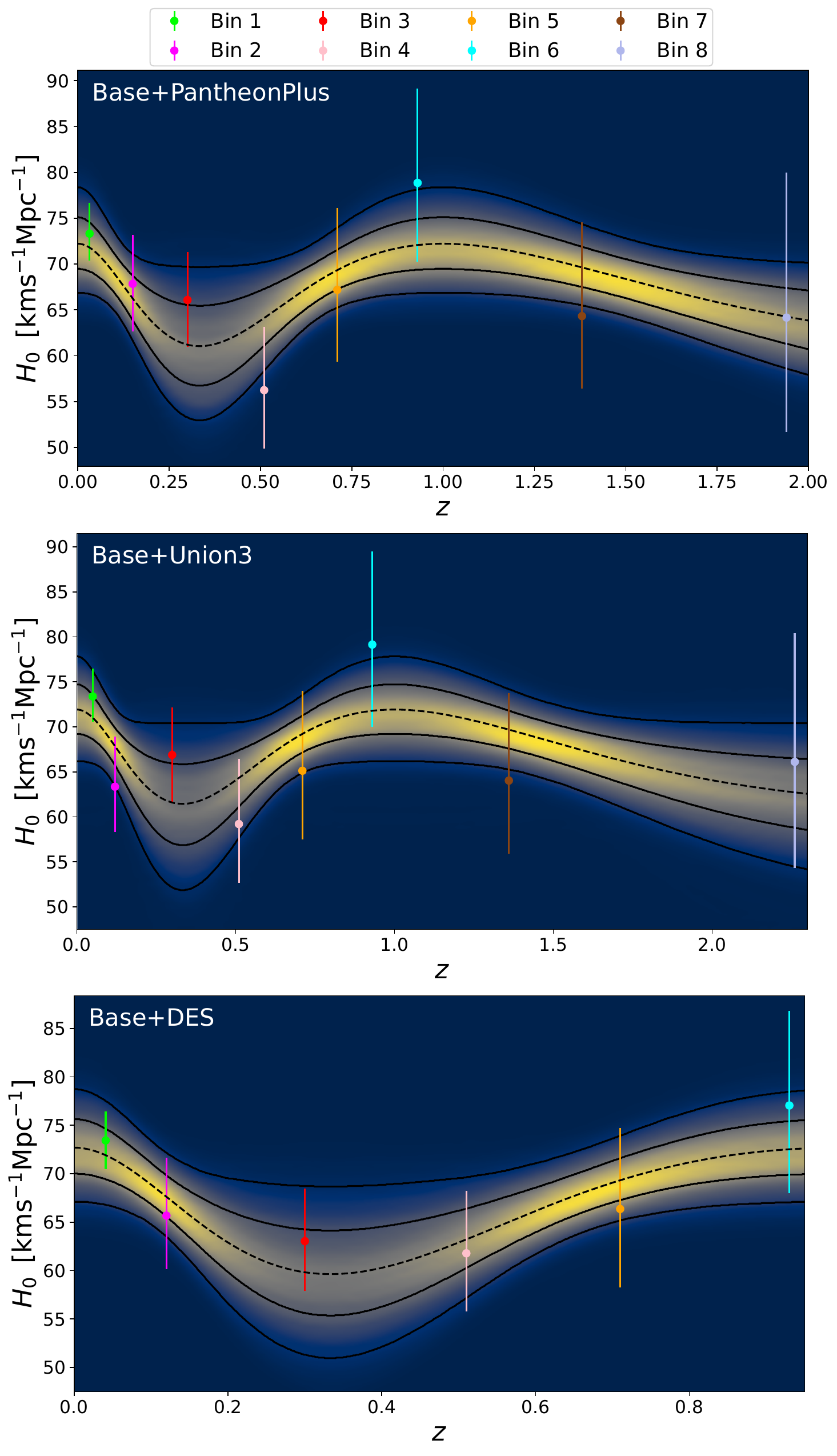}
\caption{\footnotesize{The fit of the Fourier parameterization \eqref{eq: modelo cos(2x)} to the binned results for $H_0$ in the Base+PantheonPlus sample, Base+Union3 sample and Base+DES sample. Recall that Base+DES only has 6 bins. The dashed line represents the best fit, while the external solid lines represent the 1$\sigma$ and 2$\sigma$ errors.}}
\label{fig:ajuste_cos2x}
\end{minipage}
\end{figure}

We now fit all the available bins showing the results in Table \ref{tab:resultados cos2x}. Figure \ref{fig:ajuste_cos2x} displays the fit for the 8 bins in the Base+PantheonPlus and Base+Union3 samples, and for only 6 bins in the Base+DES sample, remembering that this dataset has a reduced redshift range.

Again, Table \ref{tab:resultados cos2x} includes the $\sigma$, $\Delta$AIC, and $B$ values to see the robustness of this model against a constant $H_0$. The parameter $A$ is different from zero up to $\sim2.3$ standard deviations, suggesting a preference for the non-constant model. However, the scale in \cite{AIC2} indicates that the difference in the AIC values is insufficient to favor any of the models. Table \ref{tab:resultados cos2x} also contains the results in each dataset for only 6 bins, confirming a general improvement over the quadratic parameterization. On Jeffreys’ scale \cite{jeffreys1998theory}, the Bayes factor accounts for a barely worth mentioning value, with the exception of the Base+DES sample, where we found a substantial preference for the non-constant model. It is interesting to note that the latest two bins in Base+PantheonPlus and Base+Union3 don't change too much the results in each dataset compared to 6 bins. 

\begin{table}
\renewcommand{\arraystretch}{1.3}
\centering
\begin{tabular}{c c c c c c }
\hline
Data & $\hat{H}_{0} \left[ \frac{\text{km}}{\text{s Mpc}} \right]$ & $A \left[ \frac{\text{km}}{\text{s Mpc}} \right]$ & $\sigma$ & $\Delta$AIC & $B$ \\ \hline
\multicolumn{6}{c}{6 bins} \\ \hline
Base+PantheonPlus & $67.15^{+2.40}_{-2.35}$ & $5.70^{+2.82}_{-2.88}$ & 2 & $-1.90$ & 2.11 \\ 
Base+Union3 & $67.10^{+2.34}_{-2.47}$ & $5.35^{+3.08}_{-2.94}$ & $1.8$ & $-1.22$ & 1.90\\
Base+DES & $66.15^{+2.39}_{-2.19}$ & $6.53^{+2.88}_{-2.85}$ & $2.3$ & $-2.90$ & 3.87\\ \hline
\multicolumn{6}{c}{8 bins} \\ \hline
Base+PantheonPlus & $66.63^{+2.39}_{-2.27}$ & $5.59^{+2.76}_{-2.78}$ & 2 & $-2.03$ & 2.21 \\ 
Base+Union3 & $66.68^{+2.25}_{-2.34}$ & $5.24^{+2.97}_{-2.91}$ & $1.8$ & $-1.26$ & 1.54\\ \hline
\end{tabular}
\caption{\footnotesize Constraints on the parameters $\hat{H}_0$ and $A$ in the Fourier parameterization \eqref{eq: modelo cos(2x)}. Remember that Base+PantheonPlus and Base+Union3 reach up to $z=2.3$ while Base+DES only up to $z=1.13$.  We compared the model to a constant $H_0$ and determine the Bayes factor $B$, $\Delta$AIC and $\sigma$ deviation. We see that the deviation from a constant is larger in $\sigma$'s than for the quadratic model. The $\Delta$AIC is slightly better for the Fourier model than for a constant, although from the scale in \cite{AIC2} there is no preference to either model. The Bayes factors values account for a barely worth mentioning value according to the Jeffreys' scale \cite{jeffreys1998theory}, with the exception of the Base+DES data, which shows a substantial preference for the non constant model.}
\label{tab:resultados cos2x}
\end{table}

\subsection{Unbinned results \label{sec:unbinned}}

\begin{figure}[H]
\centering
\includegraphics[scale=0.42]{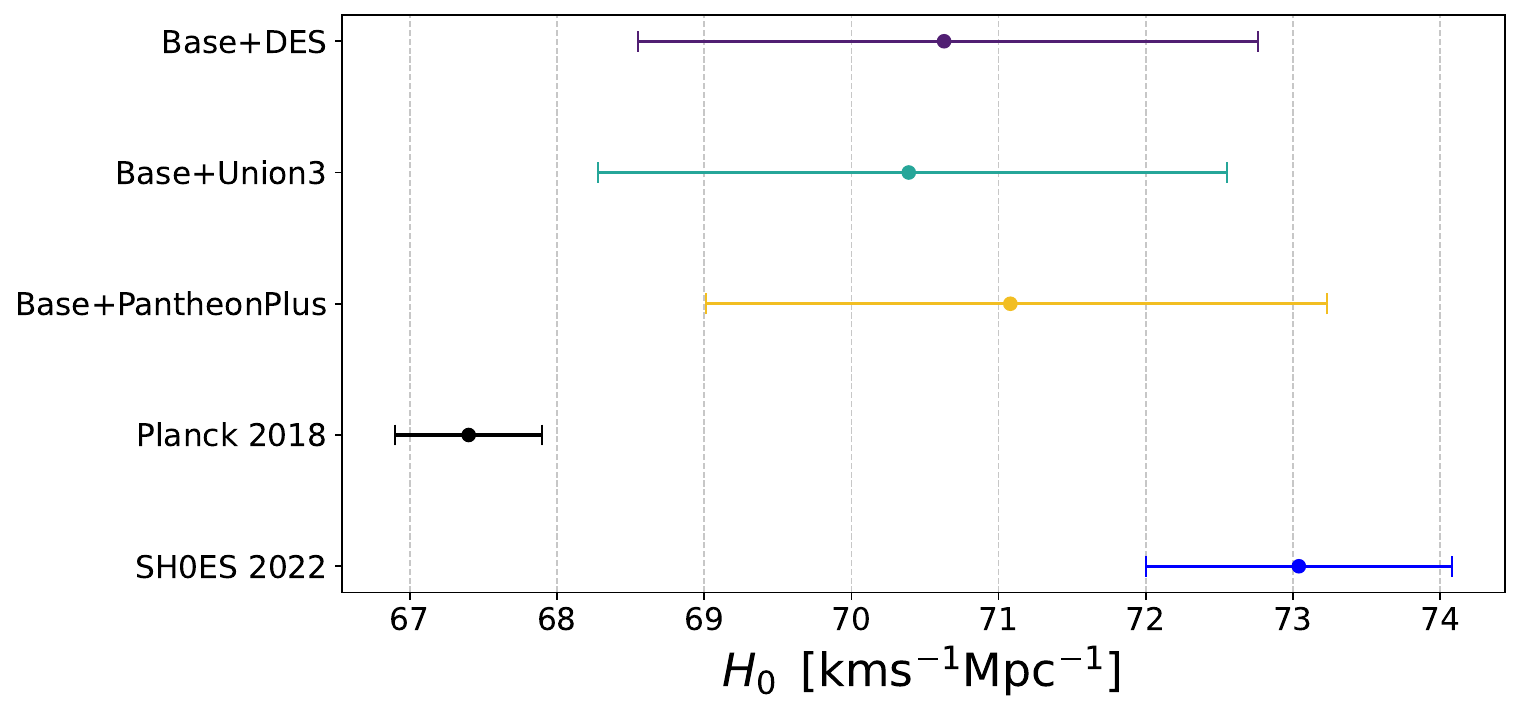}
\caption{\footnotesize{Estimates of the Hubble constant $H_0$ for the unbinned dataset Base+PantheonPlus, Base+Union3 and Base+DES, compared to $H_0$ estimates for SH0ES \cite{Riess_2022} ($H_0=73.04 \pm 1.04$ km $\text{s}^{-1}$$ \text{Mpc}^{-1}$) and Planck 2018 team \cite{refId0} ($H_0=67.4 \pm 0.5$ km $\text{s}^{-1}$$ \text{Mpc}^{-1}$).}}
\label{fig: Comparación de H0}
\end{figure}

In Table \ref{tab:resultados parametros sin binear} we show the fitted values of the cosmological parameters using the full redshift range of the data. For the Hubble constant, we obtain an intermediate value between CMB and SH0ES of $H_0 \approx 70$ km $\text{s}^{-1}$$ \text{Mpc}^{-1}$, see Figure \ref{fig: Comparación de H0}. The constraints in $w_0$ and $w_a$ are in good agreement with those found for DESI+SNe in \cite{DESI}. Table \ref{tab:resultados parametros sin binear} shows that a cosmological constant $\Lambda$ (corresponding to $w_0=-1$ and $w_a=0$) as a source of DE is excluded at approximately $2 \sigma$ with Base+Union3 and Base+DES data, while Base+PantheonPlus excludes it at approximately $1 \sigma$. Moreover, $\Omega_m$ agrees with the Planck value, $\Omega_m=0.315 \pm 0.007$; however, the sound horizon $r_d$ disagrees with the Planck value, $r_d=147.09 \pm 0.26$, by approximately $1.5\sigma$. 

\begin{table*}[t]
    \renewcommand{\arraystretch}{1.3}
    \centering
    \begin{tabular}{c c c c c c}
    \hline
    Data & ${H}_{0} \left[ \frac{\text{km}}{\text{s Mpc}} \right]$ & $\Omega_m$ & $w_0$ & $w_a$ & $r_d$ [Mpc] \\ \hline
    Base+PantheonPlus & $71.08^{+2.15}_{-2.07}$ & $0.31 \pm 0.02$ & $-0.88^{+0.08}_{-0.07}$ & $-0.45^{+0.58}_{-0.61}$ & $140.95^{+4.29}_{-4.11}$ \\ 
    Base+Union3 & $70.39^{+2.16}_{-2.11}$ & $0.33 \pm 0.02$ & $-0.68 \pm 0.13$ & $-1.35^{+0.74}_{-0.78}$ & $139.58^{+4.31}_{-4.13}$ \\ 
    Base+DES & $70.63^{+2.13}_{-2.08}$ & $0.33 \pm 0.02$ & $-0.75 \pm 0.09$ & $-1.17^{+0.64}_{-0.66}$ & $140.13^{+4.26}_{-4.20}$ \\ \hline
    \end{tabular}
    \caption{\footnotesize{Best-fit values on the parameters $H_0$, $\Omega_m$, $w_0$, $w_a$ and $r_d$ of the megamasers+CC+DESI+SNe
    dataset, without binning and over the entire redshift range $z<2.33$. Without the CC and megamasers data, $H_0$, $r_d$ and M cannot be determined. We see that the constraints on $w_0$ and $w_a$ resemble those in \cite{DESI}, for DESI + SNe data. A cosmological constant ($w_0=-1$ and $w_a=0$) is excluded by $\sim 2\sigma$ when the SNe data are either Union3 or DES. } }
    \label{tab:resultados parametros sin binear}
\end{table*}

\subsection{\texorpdfstring{$H(z)$}{H(z)} reconstruction}
Finally, given this apparent variability in $H_0$, we wonder about deviations in the Hubble parameter, $H(z)$, from its theoretical prediction in the context of the CPL model. To do so, we will treat $H(z)$ as a free parameter to be constraint directly from the full data. Specifically, $H(z)$ will be determined by its values in a set of redshift nodes $H(z_i)$, defined by equally spaced values of the scale factor $a_i$, starting at $a_1=1$. This choice is motivated by the fact that most of the data are in the range $z<1$, making it convenient for most nodes to be distributed within this region. The maximum redshift node will also be determined by the maximum of the previous bins in Table \ref{tab: bines}, i.e. $z_{\textrm{max}}=2.3$ for Base+PantheonPlus and Base+Union3, and $z_{\textrm{max}}=1.13$ for Base+DES, in the latter we used only the data with $z \leq 1.13$. We used 9 nodes for Base+PantheonPlus and Base+Union3, and 7 nodes for Base+DES.

\begin{figure}[t]
\centering
\includegraphics[scale=0.43]{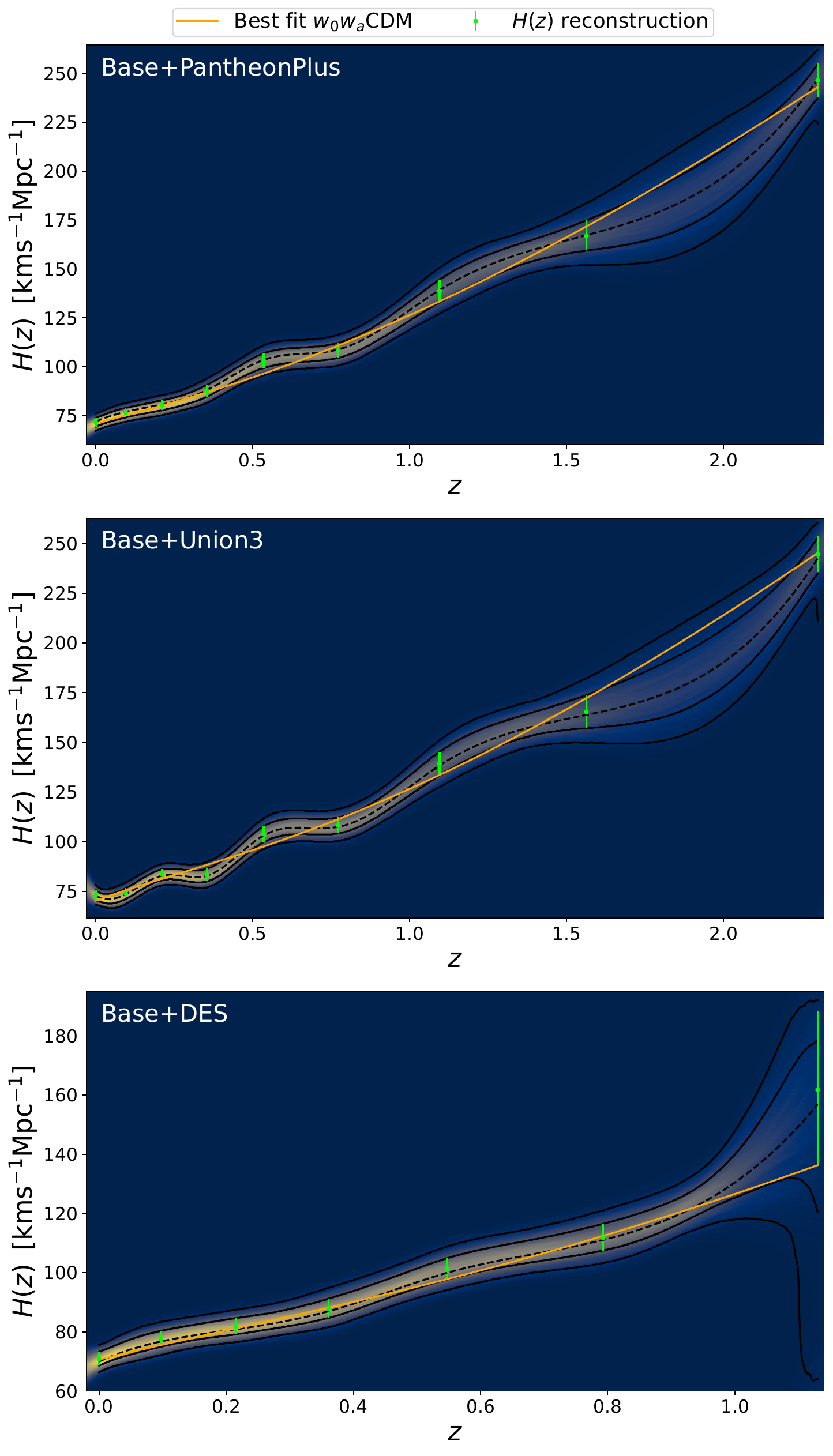}
\caption{\footnotesize{Reconstructed Hubble parameter $H(z)$ in each sample. The dashed line represents the interpolation obtained interpolating with a cubic spline the values of $H(z_i)$ in each node, while the external black solid lines represent the $1\sigma$ and $2\sigma$ errors. The orange line shows the theoretical Hubble parameter following the CPL model, with the parameters values fixed from Table \ref{tab:resultados parametros sin binear}}.}
\label{fig: H(z) spline}
\end{figure}

Each of the nodes $H_i=H(z_i)$ is treated as a free parameter, instead of the cosmological parameters $\Omega_m$, $\omega_0$ and $\omega_a$. The Hubble parameter at any $z$ interpolated with a cubic spline. Figure \ref{fig: H(z) spline} shows the reconstruction and Table \ref{tab:resultados spline} contains the Hubble constant (first node), and compares the fitting of this model compared with the CPL model.

Base+DES has large error bars in its last node, due to the quality of data in this range. This has an immediate effect on the values of $\sigma$ and $\Delta$AIC, showing a preference for the CPL model. In contrast, using Base+PantheonPlus and Base+Union3, there is a preference for the reconstructed Hubble parameter over the CPL model, with Base+Union3 being the most notable, reaching a statistical significance of up to $3.5\sigma$ and a $\Delta$AIC value indicating considerable support according to the scale in \cite{AIC2} compared to the CPL model. In order to interpret the results, it is important to note the impact of the interpolation method used. Cubic splines can present overfitting in regions where the data is disperse, generating oscillations that could be interpreted as real features of the model, but are actually mathematical artifacts. However, the values of $\Delta$AIC were not significantly affected when we used quadratic or linear splines. This shows that the choice of the interpolation scheme does not have a significant effect on the reconstruction of the Hubble parameter in the range where the data lie.

Paying attention only to the inferred nodes $H(z_i)$, we observe a slight tension around $z \sim 0.5$ for Base+PantheonPlus and Base+DES, whereas the others agree with its theoretical prediction.

\begin{table}[t]
    \renewcommand{\arraystretch}{1.7}
    \centering
    \begin{tabular}{c c c c }
    \hline
    Data & $H_{0} \left[ \frac{\text{km}}{\text{s Mpc}} \right]$ & $\sigma$ & $\Delta$AIC \\ \hline
    Base+PantheonPlus & $71.74^{+2.15}_{-2.30}$ & 3.0 & -1.91 \\ 
    Base+Union3 & $73.28^{+2.53}_{-2.43}$ & 3.5 & -5.26 \\ 
    Base+DES & $70.85^{+2.42}_{-2.24}$ & 1.2 & 3.04 \\ \hline
    \end{tabular}
    \caption{\footnotesize{The Hubble constant, $\sigma$ deviation, and $\Delta$AIC values between the reconstructed and the theoretical Hubble parameter. Base+DES favors the CPL model, probably due to the large error bars in the last node. For Base+PantheonPlus and Base+Union3, there is a preference for the reconstructed $H(z)$, as indicated by the $\sigma$ values, which rise up to $3.5\sigma$. The $\Delta$AIC values show no preference for either model from the scale in \cite{AIC2} for Base+PantheonPlus, but Base+Union3 shows considerable support against the theoretical predictions.}}
    \label{tab:resultados spline}
\end{table}

\section{\label{sec:conclu}Summary and conclusions}

In this work, we have studied the evolution in the value of $H_0$ with the redshift of the data used to determine it, finding a dynamic evolution in $H_0$. For that we used the latest dataset of CC, megamaser, SNe Ia and BAO observations from the Year 1 data release of DESI. The results show that the Hubble tension, referring to a phenomenologically non-constant value in $H_0$, cannot be resolved by the new DESI data together with a dynamical DE.

SNe observations are chosen from one of the three most recent samples of Pantheon+, Union3, and DES. Pantheon+ and Union 3 share about 1360 SNe, but differ in their analysis and treatment over systematics. DES has 1635 new SNe but shares 194 low-redshift SNe with Pantheon+. Our three data samples share the megamasers+CC+DESI data, but differ in the SNe data. Base+PantheonPlus and Base+Union3 formed a total of 8 bins, and 6 bins for Base+DES.

If we consider only the first six bins, we found (see Fig. \ref{fig:all_results_H0}), for the three samples, that $H_0$ decreases as a function of redshift for $z<0.5$, and then increases in the region $0.5<z<0.9$  up to $\sim 78$ km $\text{s}^{-1}$$ \text{Mpc}^{-1}$. In fact, if we include bins 7 and 8, the dynamic trend becomes more robust, with $H_0$ decreasing once again for these bins.

The results show a dynamical trend in $H_0$ with a statistical significance up to $1.5\sigma$, for the quadratic parameterization \eqref{eq: Función cuadrática} and up to $2.3\sigma$ for the Fourier parameterization \eqref{eq: modelo cos(2x)}, indicating a preference for a non-constant model. However, the values in $\Delta$AIC and the Bayes factor $B$, which penalize the extra parameters in the dynamical models, do not provide substantial support for the quadratic function against a constant value. From a Bayesian point of view, both the quadratic function and a constant $H_0$ fit the bins results equally well. Similar conclusions are obtained for the Fourier parameterization, except for the results of Base+DES, which have the highest $\Delta$AIC and $B$ values showing a substantial preference for the dynamical model, according to Jeffrey's scale.

Remarkably, the value of $\widetilde{H}_0 (z=0)$ for both models, as shown in Tables \ref{tab:resultados cuadrática} and \ref{tab:resultados cos2x}, is in complete agreement with that obtained by SH0ES, $H_0 = 73.04 \pm 1.04$ km $\text{s}^{-1}$$ \text{Mpc}^{-1}$.

We highlight that the three datasets, which use different SNe samples, show similar behaviors even though each sample differs in its content and methodology. Because of this, we can infer that this trend is not caused by uncountable systematic errors in the SNe parameters, and it is more an intrinsic behavior guided by the data.

We have ignored any calibration on either $r_d$ or M. However, if we adopt them, we find that the bins values of $H_0$ are completely consistent with a constant. This would require using data outside of the redshift bin, either CMB data to determine $r_d$ or distance leader data to determine M. As fixing  $r_d$  or  M  is equivalent to fixing  $H_0$  from the beginning, which prevents the data from speaking for itself. It is relevant to highlight that any solution to the Hubble tension that only modifies the calibration of $r_d$ or M will not solve the tension presented here. In the unbinned case, in subsection \ref{sec:unbinned} we obtained a $r_d$ that is $1.5\sigma$ smaller than the one measured by Planck. 

Of course, the results for $H_0$ may vary if one adopts a different binning strategy, but we expect that the significance of the trend does not change significantly. This evolution in the Hubble constant might offer a new perspective on the Hubble tension. Since we are using the $w_0 w_a$CDM model with very broad priors, it seems that a dynamical DE cannot remove this trend. It's important to note that using a cosmological constant as the Dark Energy component only reduces the errorbars in the estimation of $H_0$ due to the reduced parameter space. This may increase the significance of the non-constant models.

Some possible solutions would be either a hidden systematic error or some other assumption in the model. This could even be the reason why SH0ES and Planck have different values for $H_0$ given that they use data with different redshifts, but this needs further study, maybe considering $H_0$ not as an integration constant, but
as a quantity dependent on the observed region of the
Universe. Unfortunately, directly corroborating that the Hubble parameter at the present time has different values depending on the region of the Universe is difficult; due to the finiteness of the speed of light, we observe those regions in the past. However, nothing prevents us from thinking that just as there are perturbations to matter, these bring with them anisotropies in the expansion rate of the Universe, with $H_0$ being more of an average value and the local departure of the cosmological principle \cite{refId0-Schwarz,Bolejko_2016,refId0-Hans,10.1093/mnras/staa2348,PhysRevD.103.L121302,Heinesen_2022,PhysRevD.105.063514,PhysRevD.106.103527,PhysRevD.108.123533,Hu_2024,refId0-Hu,10.1093/mnras/stad3357}. See \cite{Kumar_Aluri_2023} for a review that surveys observational hints for deviations of the cosmological principle.

We performed a reconstruction of the Hubble parameter, $H(z)$, as an alternative strategy to binning the data, treating it as a free function determined by its values at specific redshift nodes, $H(z_i)$, which were used as fitting parameters. We then interpolated these values using cubic splines to obtain a continuous function for $H(z)$. Our analysis of the reconstructed Hubble parameter reveals deviations from the theoretical prediction under the CPL model, with Base+Union3 showing the most significant preference for the reconstructed $H(z)$, reaching a statistical significance of up to $3.5\sigma$ and a $\Delta$AIC value that indicates considerable support. Additionally, a slight tension is observed around $z \sim 0.5$ in Base+PantheonPlus and Base+DES between the reconstructed model and the CPL predictions. While checking different interpolation methods, including quadratic and linear splines, did not significantly our results.

Our binning strategy splits the data used, reducing the available data in each bin. This, combined with the decreasing quality in the data, produces large error bars in the inferred parameters which get worse at large $z$'s.
Therefore, the quadratic and Fourier fittings present large uncertainties.  We do not rule out that the constant $H_0$ is the correct result; we need to wait to improve the quality and number of the data to be able to reach a final conclusion.

The Hubble tension remains one of the most significant challenges in modern cosmology and astrophysics, highlighting persistent discrepancies in $H_0$ values derived from different observational methods and across various redshifts. This work contributes to the present discussion, providing further evidence that the Hubble tension is not only confined to early- versus late-time measurements but also appears consistently within purely late-time measurements reflecting unresolved aspect of cosmological models and data calibration.

\acknowledgments

The authors thankfully acknowledge Sebastien Fromenteau for his useful comments in the development of this work. M. L. H. was supported by Secihti grant 806098.

\bibliographystyle{JHEP}
\bibliography{bib.bib}

\end{document}